\documentclass[10pt,conference]{IEEEtran}
\IEEEoverridecommandlockouts

\usepackage{cite}
\usepackage{amsmath,amssymb,amsfonts}
\usepackage{graphicx}
\usepackage{textcomp}

\def\BibTeX{{\rm B\kern-.05em{\sc i\kern-.025em b}\kern-.08em
    T\kern-.1667em\lower.7ex\hbox{E}\kern-.125emX}}

\usepackage{tabularray}
\usepackage[noend]{algpseudocode}
\usepackage{multirow}
\usepackage{xurl}
\usepackage{booktabs} 
\usepackage{amsmath}
\usepackage{graphicx}
\usepackage[belowskip=-12pt]{caption}
\usepackage{subcaption}
\usepackage{rotating}
\usepackage{tablefootnote}
\usepackage{adjustbox}
\usepackage{array}
 \usepackage{float}
\usepackage{tabularx}
\usepackage{listings}
\usepackage{color, colortbl}
\usepackage{balance} 
\usepackage{lscape}
\usepackage{xspace}
\usepackage{framed}
\usepackage{syntax}
\usepackage{parcolumns}
\usepackage{pifont}
\usepackage{soul}
\usepackage[tikz]{bclogo}
\usepackage{enumitem}
\usepackage{soul}
\usepackage{makecell}
\usepackage[many]{tcolorbox}
\usepackage{hyperref}
\usepackage{cleveref}
\usepackage{tikzsymbols}
\usepackage{textcomp}
\usepackage{wrapfig}
\usepackage{tikz}

\crefformat{section}{\S#2#1#3} 
\crefformat{subsection}{\S#2#1#3}
\crefformat{subsubsection}{\S#2#1#3}

\definecolor{maroon}{cmyk}{0,0.87,0.68,0.00}
\definecolor{mygreen}{cmyk}{0.41,0.1,0.5,0}

\newcolumntype{R}[2]{%
    >{\adjustbox{angle=#1,lap=\width-(#2)}\bgroup}%
    l%
    <{\egroup}%
}

\usepackage[linesnumbered,ruled,vlined]{algorithm2e}
\SetKwInput{KwInputs}{Inputs}                
\SetKwInput{KwOutputs}{Outputs}              
\SetKwInput{KwOutput}{Output}              

\definecolor{patchgreen}{HTML}{ccffcc}
\definecolor{patchred}{HTML}{ffcccc}
\definecolor{myyellow}{HTML}{FFFFCC}
\definecolor{mygray}{HTML}{F0F0F0}

\newcommand\printpercent[2]{\the\numexpr#1*100/#2\%}

\preto\tabular{\setcounter{rownumbers}{0}}
\newcounter{rownumbers}

\definecolor{problemblue}{RGB}{100,134,158}
\definecolor{idiomsgreen}{RGB}{0,162,0}
\definecolor{exercisebgblue}{rgb}{0,  .69,  .941}
\definecolor{deepgreen}{rgb}{0.0, 0.5, 0.0}
\definecolor{codegreen}{rgb}{0,0.6,0}
\definecolor{codegray}{rgb}{0.5,0.5,0.5}
\definecolor{codepurple}{rgb}{0.58,0,0.82}
\definecolor{backcolour}{rgb}{0.95,0.95,0.92}
\definecolor{redColor}{RGB}{255,0,0}
\definecolor{Gray}{gray}{0.1}
\definecolor{CadetBlue}{RGB}{243.0,42.1,134.0}

\newtcbtheorem[]{Theorem}{}{
        enhanced,
        sharp corners,
        colback=white,
        colframe=exercisebgblue
    }{thm}

\lstdefinestyle{code}{
  backgroundcolor=\color{gray!4},
  commentstyle=\color{codegray},
  keywordstyle=\color{codepurple},
  numberstyle=\tiny\color{codegray},
  stringstyle=\color{codegray},
  basicstyle=\ttfamily\footnotesize,
  breakatwhitespace=false,         
  breaklines=true,                 
  captionpos=b,                    
  keepspaces=true,                 
  numbers=left,                    
  numbersep=5pt,                  
  showspaces=false,                
  showstringspaces=false,
  showtabs=false,                  
  tabsize=2,
}
\lstdefinelanguage{test}{%
	language     = python,
	breaklines = true,backgroundcolor=\color{white},escapechar=!,rulecolor=\color{black}, breaklines=true,sensitive=true,  numbersep=5pt, xleftmargin=.015\textwidth, frame=tb,label=test
}
\lstdefinelanguage{source}{%
	language     = python,
	breaklines = true,
firstnumber=0,numberfirstline=false,columns=fullflexible,numbers=left,backgroundcolor=\color{white},
    rulecolor=\color{black}, 
    breaklines=true,sensitive=true, numbersep=5pt, xleftmargin=.015\textwidth, label=test
}
\newcommand*\Suppressnumber{%
  \lst@AddToHook{OnNewLine}{%
    \let\thelstnumber\relax%
     \advance\c@lstnumber-\@ne\relax%
    }%
}

\newcommand*\Reactivatenumber{%
  \lst@AddToHook{OnNewLine}{%
   \let\thelstnumber\origthelstnumber%
   \advance\c@lstnumber\@ne\relax}%
}

\definecolor{diffstart}{named}{codegreen}
\definecolor{diffincl}{named}{redColor}

\DeclareCaptionFormat{listing}{#3}
\graphicspath{{./figures/}}

\newcounter{NumObservations}
\stepcounter{NumObservations}
\definecolor{shadecolor}{rgb}{.9,.9,.9}

\definecolor{msftBlue}{RGB}{0,164,239}
\definecolor{msftGreen}{RGB}{127,186,0}
\definecolor{msftYello}{RGB}{255,185,0}
\usepackage{bigstrut}

\definecolor{vlcolor}{rgb}{0.9,0.1,0.1}

\begin{document}

\title{A Preliminary Study of Fixed Flaky Tests in Rust Projects on GitHub}

\author{
\IEEEauthorblockN{Tom Schroeder}
\IEEEauthorblockA{\small{\textit{University of Illinois Urbana-Champaign}} \\
Urbana, IL, USA \\
jts13@illinois.edu}
\and
\IEEEauthorblockN{Minh Phan}
\IEEEauthorblockA{\small{\textit{University of Illinois Urbana-Champaign}} \\
Urbana, IL, USA \\
minhnp2@illinois.edu}
\and
\IEEEauthorblockN{Yang Chen}
\IEEEauthorblockA{\small{\textit{University of Illinois Urbana-Champaign}} \\
Urbana, IL, USA \\
yangc9@illinois.edu}
}


\maketitle

\begin{abstract}
Prior research has extensively studied flaky tests in various domains, such as web applications, mobile applications, and other open-source projects in a range of multiple programming languages, including Java, JavaScript, Python, Ruby, and more.
However, little attention has been given to flaky tests in Rust—an emerging popular language known for its safety features relative to C/C++. Rust incorporates interesting features that make it easy to detect some flaky tests, e.g., the Rust standard library randomizes the order of elements in hash tables, effectively exposing implementation-dependent flakiness.
However, Rust still has several sources of nondeterminism that can lead to flaky tests.

We present our work-in-progress on studying flaky tests in Rust projects on GitHub. Searching through the closed GitHub issues and pull requests, we identified 1,146 issues potentially related to Rust flaky tests.
We focus on flaky tests that are fixed, not just reported, as the fixes can offer valuable information on root causes, manifestation characteristics, and strategies of fixes.
By far, we have inspected 53 tests. Our initial findings indicate that the predominant root causes include asynchronous wait (33.9\%), concurrency issues (24.5\%), logic errors (9.4\%), and network-related problems (9.4\%). Our artifact is publicly available at ~\cite{artifact}.

\end{abstract}


\section{Introduction}
\label{sec:introduction}
Regression testing is crucial to ensure software reliability by detecting potential bugs after code changes.
However, flaky tests can non-deterministically pass or fail when run on the same code version, significantly impacting the quality of regression test suites~\cite{luo2014empirical}. 
Common flaky tests can be categorized into order-dependent (OD) and non-order-dependent (NOD) tests.
OD tests can be flaky due to dependencies within the test suite—passing if run before specific tests and failing if run after.
NOD tests, on the other hand, may become flaky for reasons other than order dependencies, such as concurrency, asynchronous waits, network, and I/O operations.  
Implementation-dependent (ID) tests are a subcategory of NOD caused by wrong assumptions of unordered collections in the implementation of tests.

Since the seminal study of flaky tests in Apache projects done by Luo et al. in 2014~\cite{luo2014empirical}, several studies have been proposed for characterizing~\cite{thorve2018empirical, gruber2021empirical, gruber2024do, owain2021survey,lam2019root,lam2020study,lam2020large,lam2020understanding,chen2023transforming, hashemi2022empirical, barbosa2022test}, detecting
and fixing
flaky tests across various programming languages and platforms, including Android, Java, JavaScript, Python, and Ruby. However, no prior work has specifically focused on flaky tests in Rust projects. Given Rust's emerging popularity and its interesting features that could introduce non-determinism, investigating flakiness in Rust projects is a crucial area for future research.

In this study, we investigate test flakiness in Rust projects. Starting by constructing a dataset of flaky tests from Rust projects on GitHub, we identified 1,146 issues potentially related to Rust flaky tests.
By further inspecting the characteristics of 53 tests, we identified nine common root causes of test flakiness. To our knowledge, we presented the first study to investigate flaky tests in Rust on Github projects. Our findings offer valuable insights that could guide future research.

\section{Methodology}
\label{sec:methodology}
We performed a two-step methodology—filtering and investigation—to construct the Rust flaky test dataset.

\subsection{Filtering}
To accurately investigate the root causes of flakiness, we focus on \textit{already fixed} flaky tests. Our approach involves leveraging the GitHub REST API to search for issues in repositories using Rust, specifically querying for the term \textit{'flaky'}. The search was narrowed to only closed issues that have linked pull requests. Finally, we create a dataset containing all issues captured on October 29, 2024. 
This process resulted in a dataset of 1,146 issues potentially related to Rust flaky tests. 


\subsection{Investigation}
Manually investigating over 1,000 issues is time-consuming. To ensure the diversity of tests in our study within a limited time, we shuffled the dataset and performed our analysis from start to finish. At the time of submission, we have manually investigated a subset of 53 tests from 49 projects of their root causes and fix strategies to conduct a deep analysis through the following process: 
We first reviewed the description provided in each issue, which may include the test failure scenario, the assertion failure, and the hypothesized source of the issue. We then reviewed the linked pull request to verify it as the fix for flakiness. Finally, we categorized the cause of flakiness and its fixing strategy. Note that we had to exclude certain cases because (1) some issues identified in our GitHub search using 'flaky' or its variations were actually not relevant to flaky tests, and (2) the issues pertained to parts of the repository not utilizing Rust, such as Python bindings or CI infrastructure.


\section{Analysis and Future Work}
\label{sec:analysis}

\subsection{Categories of Flakiness Root Causes}

Table~\ref{root-cause} presents the categories identified for the 53 tests through manual analysis in our study. We found that 33.9\% of these tests are affected by asynchronous waits. Among these, 38\% specifically are related to issues of \textit{improper wait}, which means tests do not properly wait for asynchronous operations to complete before moving on to assertions.
Flakiness due to concurrency, logic, and network issues also appear frequently among the top causes. Additional root causes include I/O operations, randomness, time issues, unordered data and environment.

\begin{table}[htbp]
\centering
\scriptsize
\caption{Root causes of flakiness }
\begin{tabular}{@{}llll@{}}
\toprule
\textbf{Root Cause (Percent)} & \textbf{Total} & \textbf{Root Cause (Percent)} & \textbf{Total} \\ \midrule
\textbf{Async Wait (33.9\%)} & 18 & \textbf{Concurrency (24.5\%)} & 13 \\
\quad Improper Wait & 8 & \quad Robustness & 4 \\
\quad Tweak Duration & 2 & \quad Retry & 2 \\
\quad Sleep & 2 & \quad Lock & 1 \\
\quad Increase Timeout & 1 & \quad Channel & 1 \\
\quad Retry & 1 & \quad Race Condition & 1 \\
\quad Configuration & 1 & \quad Non-deterministic Environment & 1 \\
\quad Deadlock & 1 & \quad Reorder Initialization & 1 \\
\quad Channel & 1 & \quad Shared State & 1 \\
\quad Unknown State & 1 & \quad Atomics & 1 \\
\textbf{Logic (9.4\%)} & 5 & \textbf{Network (9.4\%)} & 5 \\
\quad Off by One & 2 & \quad Certs & 1 \\
\quad Inverted Conditional & 1 & \quad Connection Loss & 1 \\
\quad Deserialization & 1 & \quad Empty Header & 1 \\
\quad Sorting & 1 & \quad Retry & 1 \\
 & & \quad Reused Resource (port) & 1 \\
\textbf{I/O (5.7\%)} & 3 & \textbf{Randomness (5.7\%)} & 3 \\
\quad Flush & 1 & \quad Ranges & 1 \\
\quad Temp Files & 1 & \quad RNG & 1 \\
\quad Resource Limit & 1 & \quad Weighting & 1 \\
\textbf{Time (3.8\%)} & 2 & \textbf{Unordered data (5.7\%)} & 3 \\
\quad Inconsistent Clock & 1 & \quad Sorting & 2 \\
\quad Off by One & 1 & \quad Non-Deterministic Environment & 1 \\
\textbf{Environment (1.9\%)} & 1 & & \\
\quad Shared libraries & 1 & & \\
\bottomrule
\end{tabular}
\label{root-cause}
\end{table}

\begin{figure}
    \centering
    \includegraphics[scale=0.65]
    {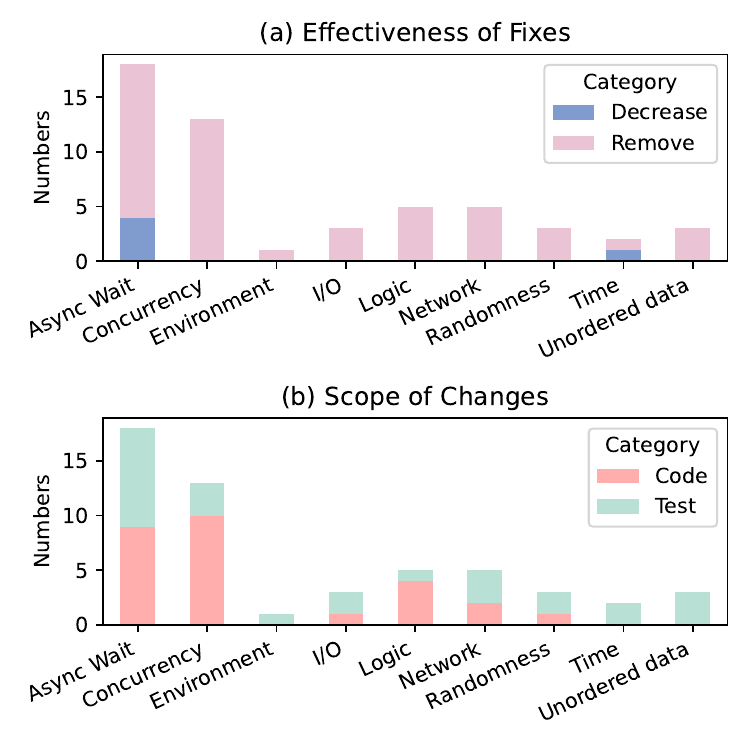}
    \vspace{-10pt}
    \caption{\small Categories of fixes. (a) shows the effectiveness of fixes per root cause category; (b) outlines the scope of changes for fixes per root cause category.}
    \label{fig:fix}
    \vspace{-4pt}
\end{figure}

\subsection{Categories of Fix Strategies}
To investigate the fixes of flaky tests, we categorize them based on the \textit{effectiveness of fixes} by assessing whether each fix completely removes flakiness or only reduces the likelihood of failure (Figure~\ref{fig:fix}(a)). Additionally, we evaluate the \textit{scope of changes}, determining whether the fix modifies test code or main code (Figure~\ref{fig:fix}(b)).
Four tests affected by async waits and one test caused by time issues have not been completely fixed but only have a reduced likelihood of test failures. The potential reason could be that the actual root cause was not fully understood or that completely removing flakiness through a patch is programmatically challenging.
Additionally, of the 53 tests, 27 were fixed by modifying the main code to address common root causes such as async waits, concurrency, and logic. The remaining tests were patched within the test code. 

\section{Future Work}
\label{sec:futurework}
We present the first study of flaky tests in Rust from GitHub projects. In future work, we aim to continue our study of the dataset with 1,146 issues related to potential Rust flaky tests and explore methods for detecting and addressing flakiness in Rust. Additionally, we will investigate how language features influence flaky tests across different programming languages.

\bibliographystyle{IEEEtran}
\bibliography{reference}

\end{document}